\documentstyle[pictex]{article}
\newcommand{\ra}{\rightarrow}

\newcommand{\be}{\begin{equation}}
\newcommand{\ee}{\end{equation}}
\newcommand{\bea}{\begin{eqnarray}}
\newcommand{\eea}{\end{eqnarray}}

\newcommand{\ov}{\overline}

\begin{document}
\title{\bf Non-factorizable Contribution to $B_K$ 
of Order $\langle G^3 \rangle$}
\author{A. E. Bergan \\ 
Institute of Physics, University of Oslo, \\
P.O. Box 1048 Blindern, N-0316 Oslo, 
Norway}
\date{18. June 1996}

\maketitle
\begin{abstract}
Within the Chiral Quark Model we have calculated the effect of 
non-factorizable $\langle G^3 \rangle$ contribution to the 
$B_K$ parameter. When all diagrams are summed, a vanishing result is obtained. 
\end{abstract}

\section{Introduction}
The low energy effects of $K^0 - \overline{K}^0$ mixing has been a 
subject of a lot of discussion in the literature. See for example  \cite{bijprades},\cite{kkmixpdr}, \cite{pradesetal} and references therein. 
Its amplitude is parametrized by the $B_K$ parameter and is conventionally defined by 
\be
\langle K^0 | {\cal O}(\Delta S =2) | \overline{K}^0 \rangle 
 = \frac{4}{3} f_K^2 m_K^2 B_K
\ee

The vacuum saturation approach accomodates only a factorizable 
contribution which yields $B_K=3/4$ in the large $N_c$ limit. By 
adding the factorizable next to 
leading order contribution $1/N_c$ increases $B_K$ to 1. 
Lattice calculations seem to indicate that the 
physical $B_K$ is numerically around $0.7$. 
Non-perturbative and non-factorizable effects can be taken into account by dressing the Feynman 
diagrams with soft external gluons in all possible ways. 
In doing this one makes for instance the identification 
$G_{\mu \nu}^a G^{a, \mu \nu} \ra  \langle G^2 \rangle$ where 
$G_{\mu \nu}$ is the gluon tensor of dimension two. 
The coefficient of the $\langle G^2 \rangle$ term has been calculated by 
Pich and de Rafael \cite{pichderaf}. 
They got a significant 
contribution to $B_K$. Numerically, their result change very much the next to leading $1/N_c$ contribution. However, due to large uncertainties in 
the experimental value of the condensate $ \langle G^2 \rangle$, $B_K$ is determined to 
lie between $0.3$ and $0.6$. The purpose of this paper is not 
to find a better numerical determination of $B_K$, 
but rather to find a more general 
expression of $B_K$ in terms of non-perturbative $\langle G^3 \rangle$ 
effects. 

In this paper we use 
the Chiral Quark Model with a constituent quark mass $M$ to calculate 
the amplitude. 
This model introduces interaction terms between quarks and mesons. This 
means that hadronic matrix elements can be calculated as loop diagrams.

When calculating fermion propagators in an external field we 
use the point gauge prescription in order to get results which 
directly contain gauge invariant quantities.

There are two independent condensates with dimension six, namely 
$\langle G^3 \rangle$ and $\langle j^2 \rangle$. Since we are working in the 
heavy quark limit, $\langle j^2 \rangle$ will be given as an expansion 
in gluon condensates of higher order and inverse proportional to the heavy 
quark mass $M$ \cite{grozinpinelis} i.e. 
\be
\langle j^2 \rangle \propto \frac{\langle G^4 \rangle}{M^2} + \cdots .
\label{j2propto}
\ee
To consider $\langle j^2 \rangle$ terms we have to consider direct 
$\langle G^4 \rangle$ terms as well. 
Therefore, our 
object of investigation is the $\langle G^3 \rangle$ terms only.

\section{The Chiral Quark Model}
In the Chiral Quark Model, chiral symmetry breaking is 
taken into account by adding an extra term $\cal{L}_{\chi}$ to the 
standard QCD Lagrangian: 
\be
 {\cal L} =  {\cal L}_{QCD} + {\cal L}_{\chi} 
\ee
where
\be
{\cal L}_{QCD} = \ov{q} (i \gamma \cdot D - {\cal M}_q) q 
 - \frac{1}{4} G^{a, \mu \nu} G^a_{\mu \nu}.   
\ee
and the ${\cal L}_{\chi}$ term is only taken into account for energy/momenta 
of order $\Lambda_{\chi}$ and below. The $q$-fields denotes only $u,d,s$ 
quarks. 
The chiral symmetry breaking scale is $\Lambda_{\chi} = 0.83 \; GeV$ and 
\be
{\cal L}_{\chi} = - M \, (\ov{q}_L \Sigma q_R + \ov{q}_R \Sigma^\dagger q_L).
\ee
The prefactor $M$ is interpreted as the constituent quark mass and is of the order $200-300 \; MeV$. The pseudoscalar 
Goldstone-octet fields $\pi^a$ are contained in 
\be
 \Sigma = exp(i \sum_a \lambda^a \pi^a / f)
\ee
In the limit where ${\cal M}_q \ra 0$, the model is $SU(3)_L \otimes SU(3)_R$ invariant. 


The chiral symmetry is thus realized through the existence 
of pseudoscalar Goldstone bosons. 

This model connects mesons and quarks and one can calculate hadronic matrix 
elements as loop diagrams. By expanding $\Sigma$ one finds the coupling 
between 
a quark density and a meson. For the $\ov{q} q meson$ vertex, the 
relevant coupling constant is $\propto M \gamma_5/f$ where $f$ is associated 
with the 
physical pion decay constant and the proportionality factor is given by 
$SU(3)$ symmetry. In the following, 
mesons are drawn by dotted lines while quarks are 
drawn with solid lines.

In the $SU(3)$ limit without gluonic condensates 
the pion decay constant is defined by 
\bea
f_{\pi}  &=&  \frac{N_c M^2}{4 \pi^2 f} {\hat f}_{\pi} \label{raw} \\
         &\propto&  "
                \setdashes <1.50pt>  
  		\plot 5 3.0 15 3.0 / 		
		\plot 5 3.2 15 3.2 /  
		\plot 5 3.4 15 3.4 /  
		\plot 5 3.6 15 3.6 /  
		\plot 5 3.8 15 3.8 / 
                \thicklines
                \setsolid
        	\put {\circle{12}} [B1] at 27 3.3 
\put {$\times$} at 30.1 3.3  
\hspace{1.3cm} " (vertex: \gamma^{\mu} L) 
\eea
With a cut-off $\Lambda$, the 
explicit expression for ${\hat f}_{\pi} = ln(\Lambda^2/M^2) + \cdots$ while in 
dimensional regularization ${\hat f}_{\pi} = (2-D/2)^{-1} + \cdots$. 
$\Lambda$ is of the order of 
the chiral symmetry breaking scale $\Lambda_{\chi}$. Although $f_{\pi}$ and $f$ appear 
differently in the Chiral Quark Model, they are for practical purposes numerically equal 
$(f_{\pi} = 93.3 \; MeV)$.

\section{Diagrammatic Calculation of $K^0 - \overline{K}^0$} 
The lowest order local operator, of dimension six, 
contributing to $K^0 - \overline{K}^0$ is 
\be
{\cal O}(\Delta S =2) = 
(\overline{d} \gamma_{\mu} L s)(\overline{d} \gamma^{\mu} L s). 
\ee
This operator 
gives rise to two diagrams pictorially given as
\be
\langle K^0 | {\cal O}(\Delta S =2) | \overline{K}^0 \rangle = 
"
                \setdashes <1.50pt>  
  		\plot 5 3.0 15 3.0 / 		
		\plot 5 3.2 15 3.2 /  
		\plot 5 3.4 15 3.4 /  
		\plot 5 3.6 15 3.6 /  
		\plot 5 3.8 15 3.8 / 
                \thicklines
                \setsolid
        	\put {\circle{12}} [B1] at 27 3.3 
                \put {\circle{12}} [B1] at 45 3.3
\put {$\times$} at 30.1 3.3
                \setdashes <1.50pt>
  		\plot 45 3.0 55 3.0 / 		
		\plot 45 3.2 55 3.2 /  
		\plot 45 3.4 55 3.4 /  
		\plot 45 3.6 55 3.6 /  
		\plot 45 3.8 55 3.8 /
\hspace{2.2cm} +   
                \setsolid
                \thicklines
\circulararc 156 degrees from 26.5 5.7 center at 21 3.3
\circulararc 156 degrees from 26.4 5.7 center at 21 3.3
\circulararc 156 degrees from 26.3 5.7 center at 21 3.3
\circulararc 156 degrees from 15.0 3.3 center at 21 3.3
\circulararc 156 degrees from 15.1 3.3 center at 21 3.3
\circulararc 156 degrees from 15.2 3.3 center at 21 3.3
\circulararc 156 degrees from 37.8 3.3 center at 32.0 3.3
\circulararc 156 degrees from 37.9 3.3 center at 32.0 3.3
\circulararc 156 degrees from 38.0 3.3 center at 32.0 3.3
\circulararc -156 degrees from 37.8 3.3 center at 32.0 3.3
\circulararc -156 degrees from 37.9 3.3 center at 32.0 3.3
\circulararc -156 degrees from 38.0 3.3 center at 32.0 3.3
\put {$\times$} at 26.6 3.3
                \setdashes <1.50pt> 
  		\plot 5 3.0 15 3.0 / 		
		\plot 5 3.2 15 3.2 /  
		\plot 5 3.4 15 3.4 /  
		\plot 5 3.6 15 3.6 /  
		\plot 5 3.8 15 3.8 /                    
  		\plot 38 3.0 48 3.0 / 		
		\plot 38 3.2 48 3.2 /  
		\plot 38 3.4 48 3.4 /  
		\plot 38 3.6 48 3.6 /  
		\plot 38 3.8 48 3.8 /   
\hspace{2.0cm} "
\label{symb1}
\ee 
The first diagram is $\propto N_c^2$ while the second diagram is 
colour suppressed and is $\propto N_c$. 

There are also other contributions than eq.\ref{symb1} with derivatives 
acting on the quantum fields. They are 
generated from the siamese penguin operator \cite{siameseoper}, \cite{joeogmeg} or similar non-local operators.   

In some calculations it is customary to Fierz transform the operator 
${\cal O}$. We can take some advantages of the transformed operator as 
some of the generated diagrams cancel. 
Consider the Fierz transformed operator:
\be
{\cal O}_F = \frac{1}{N_c} (\overline{d} \gamma_{\mu} L s)(\overline{d} \gamma^{\mu} L s) + 2 (\overline{d} \gamma_{\mu} L t^a s)(\overline{d} \gamma^{\mu} L t^a s) 
\ee 
The matrix element can now be written symbolically:
\bea
\langle K^0 | {\cal O}_F | \overline{K}^0 \rangle &=& 
\frac{1}{N_c}  \left\{ " 
                \setdashes <1.50pt>  
  		\plot 5 3.0 15 3.0 / 		
		\plot 5 3.2 15 3.2 /  
		\plot 5 3.4 15 3.4 /  
		\plot 5 3.6 15 3.6 /  
		\plot 5 3.8 15 3.8 / 
                \thicklines
                \setsolid
        	\put {\circle{12}} [B1] at 27 3.3 
                \put {\circle{12}} [B1] at 45 3.3
\put {$\times$} at 30.1 3.3
                \setdashes <1.50pt>
  		\plot 45 3.0 55 3.0 / 		
		\plot 45 3.2 55 3.2 /  
		\plot 45 3.4 55 3.4 /  
		\plot 45 3.6 55 3.6 /  
		\plot 45 3.8 55 3.8 /
\hspace{2.2cm} +   
                \setsolid
                \thicklines
\circulararc 156 degrees from 26.5 5.7 center at 21 3.3
\circulararc 156 degrees from 26.4 5.7 center at 21 3.3
\circulararc 156 degrees from 26.3 5.7 center at 21 3.3
\circulararc 156 degrees from 15.0 3.3 center at 21 3.3
\circulararc 156 degrees from 15.1 3.3 center at 21 3.3
\circulararc 156 degrees from 15.2 3.3 center at 21 3.3
\circulararc 156 degrees from 37.8 3.3 center at 32.0 3.3
\circulararc 156 degrees from 37.9 3.3 center at 32.0 3.3
\circulararc 156 degrees from 38.0 3.3 center at 32.0 3.3
\circulararc -156 degrees from 37.8 3.3 center at 32.0 3.3
\circulararc -156 degrees from 37.9 3.3 center at 32.0 3.3
\circulararc -156 degrees from 38.0 3.3 center at 32.0 3.3
\put {$\times$} at 26.6 3.3
                \setdashes <1.50pt> 
  		\plot 5 3.0 15 3.0 / 		
		\plot 5 3.2 15 3.2 /  
		\plot 5 3.4 15 3.4 /  
		\plot 5 3.6 15 3.6 /  
		\plot 5 3.8 15 3.8 /                    
  		\plot 38 3.0 48 3.0 / 		
		\plot 38 3.2 48 3.2 /  
		\plot 38 3.4 48 3.4 /  
		\plot 38 3.6 48 3.6 /  
		\plot 38 3.8 48 3.8 /   
\hspace{2.0cm} " \right\}  (vertex: \gamma^{\mu} L) \nonumber \\ 
 & & + 2  \left\{ "  
                \setdashes <1.50pt>  
  		\plot 5 3.0 15 3.0 / 		
		\plot 5 3.2 15 3.2 /  
		\plot 5 3.4 15 3.4 /  
		\plot 5 3.6 15 3.6 /  
		\plot 5 3.8 15 3.8 / 
                \thicklines
                \setsolid
        	\put {\circle{12}} [B1] at 27 3.3 
                \put {\circle{12}} [B1] at 45 3.3
\put {$\times$} at 30.1 3.3
                \setdashes <1.50pt>
  		\plot 45 3.0 55 3.0 / 		
		\plot 45 3.2 55 3.2 /  
		\plot 45 3.4 55 3.4 /  
		\plot 45 3.6 55 3.6 /  
		\plot 45 3.8 55 3.8 /
\hspace{2.2cm} +   
                \setsolid
                \thicklines
\circulararc 156 degrees from 26.5 5.7 center at 21 3.3
\circulararc 156 degrees from 26.4 5.7 center at 21 3.3
\circulararc 156 degrees from 26.3 5.7 center at 21 3.3
\circulararc 156 degrees from 15.0 3.3 center at 21 3.3
\circulararc 156 degrees from 15.1 3.3 center at 21 3.3
\circulararc 156 degrees from 15.2 3.3 center at 21 3.3
\circulararc 156 degrees from 37.8 3.3 center at 32.0 3.3
\circulararc 156 degrees from 37.9 3.3 center at 32.0 3.3
\circulararc 156 degrees from 38.0 3.3 center at 32.0 3.3
\circulararc -156 degrees from 37.8 3.3 center at 32.0 3.3
\circulararc -156 degrees from 37.9 3.3 center at 32.0 3.3
\circulararc -156 degrees from 38.0 3.3 center at 32.0 3.3
\put {$\times$} at 26.6 3.3
                \setdashes <1.50pt> 
  		\plot 5 3.0 15 3.0 / 		
		\plot 5 3.2 15 3.2 /  
		\plot 5 3.4 15 3.4 /  
		\plot 5 3.6 15 3.6 /  
		\plot 5 3.8 15 3.8 /                    
  		\plot 38 3.0 48 3.0 / 		
		\plot 38 3.2 48 3.2 /  
		\plot 38 3.4 48 3.4 /  
		\plot 38 3.6 48 3.6 /  
		\plot 38 3.8 48 3.8 /   
\hspace{2.0cm} " \right\}  (vertex: \gamma^{\mu} L t^a) \; \; \; \; \;
\label{symb2}
\eea
The Fierz transformed operator gives rise to four diagrams. 
The symbolic descriptions of ${\cal O}$ and ${\cal O}_F$ 
are of course identical. The first term in eq.\ref{symb1} corresponds 
to the second {\it and} the fourth term 
of eq.\ref{symb2}. Similarly, the second term of eq.\ref{symb1} corresponds 
to the first {\it and} the third term of eq.\ref{symb2}. Examples below show 
that this is indeed the case whether effects of gluon condensates are 
included or not.
One can in fact choose to 
consider only "factorizable" contributions with 
different vertices and prefactors (first and third term of eq.\ref{symb2} and first term of eq.\ref{symb1}). 

If the external field is turned off, one sees from the colour structure that 
the first term in eq.\ref{symb1} corresponds to the second and fourth term in eq.\ref{symb2}: 
\be
 N_c^2 = \frac{1}{N_c} N_c + 2 Tr(t^a t^a).
\ee 
Similarly, the second term in eq.\ref{symb1} corresponds to 
the first term (and a vanishing third term) in eq.\ref{symb2}. 

In order to calculate non-perturbative contributions to $B_K$, we need 
fermion propagators in an external gluon field. A pedagogical 
presentation for this calculation is presented by Novikov 
et al. in \cite{pointgauge}. 
In the point gauge the gluon field can be expanded in terms of the gluon 
tensor $G_{\beta \alpha}$: 
\be
A_{\alpha} (x) = \frac{1}{2} x^{\beta} G_{\beta \alpha} (0) + \frac{1}{3} x^{\mu} x^{\beta} [D_{\mu} G_{\beta \alpha}] (0) + \frac{1}{8} x^{\mu} x^{\nu} x^{\beta} [D_{\mu} D_{\nu}G_{\beta \alpha}] (0) + \cdots .
\label{afeltexp}
\ee
When calculating the fermion propagator in an external gluon field 
we see that we get contributions 
from "G", "DG" and "DDG" terms.
In order to obtain gluon condensate contributions one has interpret 
products of gluon tensors as vacuum averages. An averaging over 
gluon indices yields for example the substitution 
\be
 G^a_{\alpha \beta} G^b_{\mu \nu} \ra \frac{\delta^{ab}}{96} 
\langle G^2 \rangle 
(g_{\alpha \mu} g_{\beta \nu} - g_{\alpha \nu} g_{\beta \nu}) .
\label{vacaverage}
\ee 
This is how non-perturbative effects are interpreted from a perturbative 
prescription. 
Similar expressions as in eq.\ref{vacaverage}, albeit more complicated, 
can be obtained with three gluons and terms including derivatives. They can be derived 
by making use of the equations of motion 
\be
 D^{\mu} G_{\mu \nu}^a = -g_s j^a_{\nu},
\ee
the commutation relation 
\be
\left[ D_{\mu}, D_{\nu} \right] = - i g_s G_{\mu \nu} ,
\ee
the Bianchi identity and translational invariance. This will enable us 
to separate the $\langle G^3 \rangle$ contribution 
from  $\langle j^2 \rangle$. We find the relations 
\bea
\langle 0|(D_{\alpha} G_{\rho \mu})^a (D^{\alpha} G^{\rho \mu})^a |0 \rangle &=& 
2 g_s^2 \langle j^2 \rangle - 2g_s \langle G^3 \rangle \\
\langle 0|(D_{\alpha} G_{\rho \mu})^a (D^{\rho} G^{\alpha \mu})^a |0 \rangle &=& 
g_s^2 \langle j^2 \rangle - g_s \langle G^3 \rangle \\
\langle 0|(D_{\alpha} G^{\alpha \mu})^a (D_{\rho} G^{\rho}_{\; \mu})^a |0 \rangle &=& 
g_s^2 \langle j^2 \rangle 
\eea
See also reference \cite{nikolaev}.

When calculating the off-diagonal $K^0 - \overline{K}^0$ amplitude with 
external gluons, one can take 
advantage of the diagrams already included in the expression for $f_{\pi}$. 
A more refined version of eq.\ref{raw} including gluon condensates 
can be written \cite{pichderaf} 
\be
f_{\pi}^2 = \frac{N_c M^2}{4 \pi^2} \left[ {\hat f}_{\pi} + 
\frac{\pi^2}{6 N_c M^4} \langle \frac{\alpha_s}{\pi} G^2 \rangle 
 + \frac{1}{360 N_c} \frac{\langle g_s^3 G^3 \rangle}{M^6}+\cdots \right]    
\ee 
where the dots denotes higher order condensates. 
During the calculation we need some useful trace expressions of the 
$SU(3)$ generators: 
\be
Tr(t^a t^b) = \frac{1}{2} \delta^{ab} ,
\ee
\be
Tr(t^a t^b t^c) = \frac{1}{4} (d^{abc}+i f^{abc})
\ee
and 
\be
Tr(t^a t^b t^c t^d) = \frac{1}{4 N_c} \delta^{ab} \delta^{cd}+ 
\frac{1}{8} (d^{abe}+i f^{abe}) (d^{cde}+i f^{cde}) .
\ee 

When dressing the "blobs" from the ${\cal O}_F$ operator with two 
external gluons ending in vacuum, one can realize 
that the two diagrams having only one 
single colour line, i.e. the second and fourth term of eq.\ref{symb2}, indeed 
cancels in the two gluon case (one gluon on each 
loop) as seen from the colour structure 
\be
 \frac{1}{N_c} Tr(t^a t^b) +2 Tr(t^a t^c t^b t^c) = 0 .
\ee 
In addition, the first term in eq.\ref{symb2} vanishes due to traceless colour matrices. 
Thus, we only have to consider the diagrams coming from the third term in 
eq.\ref{symb2} with external gluons attached. This is a non-factorizable contribution 
due to the vertex structure $\gamma^{\mu} L t^a$. 
Non-perturbative gluons to order $\langle \frac{\alpha_s}{\pi} G^2 \rangle$ 
contributes to $B_K$ as 
\be
 B_K = \frac{3}{4} \left\{1+ \frac{1}{N_c} \left[ 1- 
\frac{ N_c \langle \frac{\alpha_s}{\pi} G^2 \rangle}{32 \pi^2 f_{\pi}^4} \right] \right\} .
\label{bkparam}
\ee 
This is in accordance with \cite{pichderaf}. 
The terms with two gluons on each loop are only contributing to $f_{\pi}$ and 
therefore bring no effect into $B_K$. One should also have in mind that 
eq.\ref{bkparam} is further modified by meson loops \cite{bruno}. 

In the case of three gluons, one again realizes that the second and fourth 
term in eq.\ref{symb2} cancels in the case of two external gluons on one loop and one gluon on the other. The cancellation is again 
seen by looking at the colour structure. 
One encounters the sum 
\be
 \frac{1}{N_c} Tr(t^a t^b t^c) + 2 Tr(t^a t^b t^d t^c t^d) = 0
\ee
where we have used the commutator and completeness relation 
of two $SU(3)$ generators and 
applied the trace expressions for up to four $SU(3)$ generators. 
The diagrams with all three gluons attached to one loop are 
already included in the physical pion decay constant. 
Thus, we are left only with the third term in eq.\ref{symb2} 
which we dress with gluons. 
Due to the different "G", "DG" and "DDG" parts in 
eq.\ref{afeltexp} there are totally 20 diagrams to calculate. 

A useful tool for the calculation is to use a software package for 
algebraic manipulation. We use the Form program, invented by 
J. Vermaseren \cite{form}, which is well suited 
for our purpose. 
By summing up all diagrams, we find zero contribution of the 
$\langle G^3 \rangle$ terms. 
This result is 
a generalization of a result obtained in the paper of 
W. Hubschmid and S. Mallik \cite{hm} 
where they show that there are no contribution of 
$\langle G^3 \rangle$ to two-point functions. 
A more complicated structure is encountered in our case 
when calculating the $B_K$ 
parameter and their result is therefore not automatically applicable in 
our case.  

As a cross-check of our result, 
we also performed the calculation with just the 
original operator ${\cal O}$. This was a more time consuming operation, but 
the same vanishing result was obtained. 

We have also calculated the contributions proportional to 
$\langle j^2 \rangle$ and obtained a non-vanishing constant 
term (constant with respect to the kaon momentum) which is not acceptable.  
However, $\langle G^4 \rangle$ type contributions should be 
included as well according to eq.\ref{j2propto}. To 
complete this part of the calculation would require a major effort. 

\section{Conclusion}
We have calculated the non-factorizable $<G^3>$ contribution to 
the $B_K$ parameter. 
A vanishing result is obtained as \cite{hm} did for the two point function. 
The next order contributions are coming from operators 
of dimension 8. 
To calculate their contribution is a quite 
extensive task. Also, due to the lack of numerical information on 
dimension 8 operators, one may wonder if that is worth while at the present 
stage. 

\section{Acknowledgements}
The author is grateful to professor A. Andrianov and professor J.O. Eeg for useful discussions and comments. 

\bibliographystyle{unsrt}  

\begin{thebibliography}{10}

\bibitem{bijprades}
J.Bijnens and J.Prades.
\newblock \newline {T}he ${B}_{K}$ {P}arameter in the $1/{N}_c$ {E}xpansion.
\newblock {\em \newline {N}uclear Physics B444}, pages 523--562, 1995.

\bibitem{kkmixpdr}
A.~Pich and E.~de~Rafael.
\newblock \newline ${K}- \overline{K}$ {M}ixing in the {S}tandard {M}odel.
\newblock {\em \newline Physics Letters B158 (6)}, pages 477--484, 1985.

\bibitem{pradesetal}
{J. Prades, C. A. Dominguez, J. A. Pe\~{n}arrocha, A. Pich and E. de Rafael}.
\newblock \newline {T}he ${K}^0- \overline{K}^0$ {B}-factor in the
  {QCD}-hadronic {D}uality {A}pproach.
\newblock {\em \newline Zeitschrift f{\"u}r Physik C51 (2)}, pages 287--295,
  1991.

\bibitem{pichderaf}
A.~Pich and E.~de~Rafael.
\newblock \newline {F}our-quark {O}perators and {N}on-leptonic {W}eak
  {T}ransitions.
\newblock {\em \newline {N}uclear Physics B358}, page 311, 1991.

\bibitem{grozinpinelis}
{A. G. Grozin and Yu. F. Pinelis}.
\newblock \newline {C}ontribution of {H}igher {G}luon {C}ondensates to
  {L}ight-{Q}uark {V}acuum {P}olarization.
\newblock {\em \newline Zeitschrift f{\"u}r Physik C33}, pages 419--425, 1987.

\bibitem{siameseoper}
{J. F. Donoghue, E. Golowich and G. Valencia}.
\newblock \newline {N}ew {F}our-quark ${\Delta} {S}=2$ {L}ocal {O}perator.
\newblock {\em \newline Physical Review D33 (5)}, page 1387, 1986.

\bibitem{joeogmeg}
A.~E. Bergan and J.~O. Eeg.
\newblock \newline ${K} - \ov{K}$ {M}ixing in a {L}ow {E}nergy {E}ffective
  {Q}{C}{D}.
\newblock {\em \newline {Z}eitschrift f{\"u}r Physik C61}, pages 511--516,
  1994.

\bibitem{pointgauge}
V.~A.~Novikov et~al.
\newblock \newline {C}alculations in {E}xternal {F}ields in {Q}uantum
  {C}hromodynamics. {T}echnical {R}eview.
\newblock {\em \newline Fortschritte der Physik 32 (11)}, pages 585--622, 1984.

\bibitem{nikolaev}
S.~N. Nikolaev and A.~V. Radyushkin.
\newblock \newline {V}acuum {C}orrections to {Q}{C}{D} {C}harmonium {S}um
  {R}ules.
\newblock {\em \newline {N}uclear {P}hysics B213}, page 285, 1983.

\bibitem{bruno}
C.Bruno.
\newblock \newline {C}hiral {C}orrections to the ${K}^0-\overline{K}^0$
  ${B}_{K}$-parameter.
\newblock {\em \newline Physics Letters B320}, pages 135--140, 1994.

\bibitem{form}
J.~A.~M. Vermaseren.
\newblock {\em \newline {Symbolic Manipulation with FORM}}, August 1989.

\bibitem{hm}
W.~Hubschmid and S.~Mallik.
\newblock \newline {O}perator {E}xpansion at {S}hort {D}istance in {QCD}.
\newblock {\em \newline Nuclear Physics B207}, pages 29--42, 1982.

\end{thebibliography}

\end{document}